# Kondo effect in a novel 5$d$ quasi-skutterudite Yb$_3$Os$_4$Ge$_{13}$

C. L. Yang,[1,2] X. Wang,[1] X. Zhang,[1] D. S. Wu,[1] M. Liu,[1,3] P. Zheng,[1] J. Y. Yao,[4] Z. Z. Li,[2] Y.-F. Yang,[1,5*] Y. G. Shi,[1*] J. L. Luo,[1,5] N. L. Wang[5,6]

[1] Beijing National Laboratory for Condensed Matter Physics & Institute of Physics, Chinese Academy of Science, Beijing 100190, China

[2] College of Physics and Information Engineering, Hebei Normal University, Shijiazhuang 050024, China

[3] College of Physical Science and Technology, Sichuan University, Chengdu 610065, China

[4] Center for Crystal Research and Development, Technical Institute of Physics and Chemistry, Chinese Academy of Sciences, Beijing 100190, China

[5] Collaborative Innovation Center of Quantum Matter, Beijing 100871, China

[6] International Center for Quantum Materials, School of Physics, Peking University, Beijing 100871, China

**Abstract**

We report the crystal growth of a new compound, Yb$_3$Os$_4$Ge$_{13}$, by using a Bi-flux method. X-ray diffraction measurement shows that it crystallizes in the quasi-skutterudite-type caged structure with a cubic space group of Pm-3n (No. 223). Magnetic measurements reveal almost fully localized Yb $f$-moments above 120 K. The resistivity exhibits a crossover from metallic to insulating behavior with a logarithmic increase below ~ 40 K. The specific heat coefficient shows a rapid upturn below ~5 K and exceeds 2 J mol$^{-1}$ K$^{-2}$ at 2 K. Our experimental analysis and electronic band structure calculations demonstrate that Yb$_3$Os$_4$Ge$_{13}$ exhibits the Kondo effect due to strong hybridization of the localized Yb $f$-moments with the $p$-electrons of the surrounding Ge-cages.

* Corresponding authors: ygshi@ iphy.ac.cn, yifeng@iphy.ac.cn

## I. Introduction

The interplay of unconventional superconductivity and magnetic/charge orders in the quasi-skutterudites $R_3T_4X_{13}$ has attracted much interest in the past decades [1-5]. Efforts in recent years have witnessed a continuous expansion of the $R_3T_4X_{13}$ family, leading to the discovery of many exotic phases. A prominent virtue of the $R_3T_4X_{13}$ family is that all its three sites ($R$, T, and X) are amenable to a wide variety of elements without altering the crystal structure, thus allowing the engineering of different magnetic exchange pathways: $R$ can be a rare-earth or an earth alkaline element, T can be a transition metal from $3d$ to $5d$, and X can be either a group-III or a group-IV element. These compounds typically crystalize into the same primitive cubic crystal structure with a space group Pm-3n, but show remarkably different properties depending on the choice of the three elements. For example, in the $R_3T_4Sn_{13}$ system, compounds with $R=$ La, Sr, Ca, Yb, Th, and T= Ir, Rh are superconductors [6-14], whereas those with $R=$ Gd and Eu order magnetically [14,15]. In the $R_3T_4Ge_{13}$ system, $R_3Ru_4Ge_{13}$ compounds with $R=$ Nd-Yb show paramagnetic behavior down to 1.5 K, while $Lu_3Ru_4Ge_{13}$ and $Y_3Ru_4Ge_{13}$ exhibit diamagnetic and Pauli paramagnetism above 4.2 K, respectively, and become superconducting at lower temperatures [3]. $Y_3Ir_4Ge_{13}$ has very large thermoelectric power and is a promising candidate for thermoelectrics [16]. The quasi-skutterudite $R_3T_4X_{13}$ compounds provide a laboratory playground for studying the physical and chemical conditions of these various correlation phenomena. Exploration of novel quasi-skutterudite $R_3T_4X_{13}$ compounds is therefore of great interest.

Among the $R_3T_4X_{13}$ family, the rare earth intermetallics are of particular interest because of their exotic properties such as the heavy fermion behavior, the Kondo effect, and other intriguing magnetic orders [15]. For example, both $Ce_3Co_4Sn_{13}$ and $Ce_3Rh_4Sn_{13}$ show heavy fermion behavior at low temperatures [17,18]. In these compounds, competition between magnetic orders and collective hybridization of localized $f$ electrons with background conduction electrons may lead to a magnetic quantum phase transition at zero temperature. New quantum states of matter such as superconductivity with unconventional non-s-wave pairing often emerge at the border of the magnetic orders. In the heavy fermion compound $Ce_3Ir_4Sn_{13}$, strange double peak feature has been observed in the specific heat near the antiferromagnetic ordering temperature [19,20]. In $Ce_3Pt_4In_{13}$, pressure experiment has revealed substantial evidences against the conventional Doniach phase diagram [21].

However, most present studies focus only on Ce-based $R_3T_4X_{13}$ compounds. Few Yb-based $R_3T_4X_{13}$ compounds have been investigated because of the high volatility of Yb during sample preparation. Among them, the germanide compounds, $X_3T_4Ge_{13}$, show remarkable and puzzling variation of properties with the change of T from 3$d$ Co to 4$d$ Rh and Ru to 5$d$ Ir element. While $Yb_3Co_4Ge_{13}$ is a metal with intermediate valence of Yb ions and has very attractive thermoelectric properties [22,23], $Yb_3Rh_4Ge_{13}$ exhibits fully localized Yb $f$-moments at high temperatures and undergoes a phase transition at 2.3 K of most probably antiferromagnetic nature. On the other hand, the high temperature effective moment of Yb ions in $Yb_3Ru_4Ge_{13}$ is only half of that estimated for a free $Yb^{3+}$ ion (4.54 $\mu_B$) [3,24]. Both show good metallic conduction. In contrast, the 5$d$ Ir-based compound, $Yb_3Ir_4Ge_{13}$, exhibits insulating behavior above 130 K and has multiple charge and magnetic orderings at low temperatures, whose physical origin remains unclear [25]. In this work, we report a new 5$d$ Os-based compound, $Yb_3Os_4Ge_{13}$, which, different from other $Yb_3T_4Ge_{13}$ family members, exhibits a crossover from metallic to insulating behavior at ~ 40 K and very large specific heat coefficient exceeding 2 J mol$^{-1}$ K$^{-2}$ at low temperatures. We will argue that these could be understood from the Kondo effect. Our discovery poses the question concerning the role of the T-elements (from 3$d$ to 5$d$) in determining the physical properties in these compounds.

**II. Experimental**

Single crystals of $Yb_3Os_4Ge_{13}$ were synthesized by a solid state reaction by using Bi as the flux. The starting materials Yb (99.99%), Os (99.99%), and Ge (99.999%) were mixed in a molar ratio of 3: 4: 13 and placed in an alumina ampoule with amount of Bi (99.99%) flux. The operation was in a glove box filled with high-purity argon. The alumina ampoule with the starting materials was sealed in a Ta tube under argon atmosphere and then in an evacuated quartz tube, followed by heating the whole assembly in a furnace at 1150 °C for 2 h. It should be noted that the temperature was increased to 1150 °C quite slowly over 10 h. After reaction, the system was cooled down to 850 °C at a rate of 2 °C/h, then to 600 °C in 20 hours. The quartz tube was then inverted and quickly spun in a centrifuge to remove the Bi flux. Crystals with approximate dimensions of 0.8×0.5×0.5 mm (see in Fig. 1a) were left in the alumina ampoule. It is apparent that the crystals have cage-like shape and show well defined (100), (110) and (101) large clean surfaces, thus allowing study on the anisotropy of the electrical and magnetic properties. The chemical composition of the crystal was analyzed by energy dispersive X-ray (EDX) spectroscopy using a Hitachi S-4800 scanning electron microscope (SEM) at an accelerating voltage of 15 KV, with an accumulation time of 90 s. The EDX measurements at different locations on crystal surface indicated that the average composition is actually stoichiometric.

Phase examination of the crystal was studied by X-ray diffraction (XRD) on a Bruker SMART APEX II diffractometer at room temperature using Mo Kα radiation (λ = 0.71073 Å). The SAINT+ and XPREP programs were used for data acquisition, extraction/reduction, and empirical absorption correction [26]. The crystal structure was refined by full-matrix least–squares fitting on F2 using the SHELXL-97 program [27]. The magnetic susceptibility ($\chi$) was measured on a Magnetic Property Measurement System (MPMS; Quantum Design) between 2 K and 300 K in various applied magnetic fields up to 50 KOe under field-cooling (FC) and zero-field-cooling (ZFC) conditions. The isothermal magnetization (M-H) was measured in a MPMS between +70 KOe and –70 KOe at 1.6 K. A large single crystal was used for the measurements. The magnetic anisotropy was also characterized on the <100> and <110> directions with the magnetic field perpendicular to the different surfaces. The magnetic susceptibility of a quartz sample holder was measured independently and subtracted from the total magnetic data. The electrical resistivity ($\rho$) and specific heat ($C_p$) were measured in PPMS. The resistivity was measured upon cooling from 300 K to 2 K using a standard four-probe technique with a gauge current of 100 mA. Platinum wires and silver paste were used to make electrical contacts on each crystal. The specific heat $C_p$ was measured by a thermal-relaxation method between 2 K and 300 K in PPMS. The electronic density of states (DOS) and band structures were calculated by the local-density approximation (LDA) method within the framework of the density functional theory (DFT) [28], using a highly precise full-potential linearized augmented-plane-wave method as implemented in the WIEN2k package [29].

**III. Results and Discussion**

Figure 1c shows the XRD patterns of the $Yb_3Os_4Ge_{13}$ single crystal measured on two different large flat surfaces. All Bragg reflections can be well indexed on the basis of the $Pr_3Rh_4Sn_{13}$-type cubic structure (space group: Pm-3n, No. 223) with the lattice parameter $a$ = 8.9302(5) Å [14]. The crystal is of high purity and the XRD data show no visible impurity peaks. The two large flat surfaces were then identified as the (100) and (110) planes. The refined crystallographic parameters are listed in Table I. The crystal structure of $Yb_3Os_4Ge_{13}$ is illustrated in Fig. 1b based on the refinement and shows a cage-type character with two 20-coordination cages surrounding the Os and Yb atoms, respectively, and a third cage centered on empty position.

The temperature dependence of the resistivity, $\rho$, of the $Yb_3Os_4Ge_{13}$ single crystal is presented in Fig. 2a. We find monotonic decrease of $\rho$ upon cooling down to 40 K and a small RRR value ($\rho$(300)

K/$\rho$(50 K) ~ 1.25). The dotted line gives our fit using the Boch-Grüneisen-Mott (BGM) model [30]:

$$\rho(T)_{BGM} = \rho_0 + 4RT\left(\frac{T}{T_D}\right)^4 \int_0^{T_D/T} \frac{x^5 dx}{(e^x-1)(1-e^{-x})} - \alpha T^3, \quad (1)$$

where $\rho_0$ is the residual resistivity due to static defects of the crystal lattice, the second term describes the scattering of conduction electron by thermally excited phonons [31], and the third term originates from the *s-d* interband scattering [30,32]. The parameter $R$ is proportional to the electron-phonon scattering magnitude and $T_D$ is the Debye temperature. We find a good fit above 50 K with $\rho_0 = 504$ μΩ cm, $R = 0.56$ μΩ cm K$^{-1}$, $\alpha = 7 \times 10^{-7}$ μΩ cm K$^{-3}$, and $T_D = 268$ K. The large value of $R$ reflects strong electron-phonon scattering in Yb$_3$Os$_4$Ge$_{13}$.

Below 40 K a rapid upturn is observed in the resistivity. As may be seen in the inset of Fig. 2a. we find this could be well explained by including a logarithmic term in the modified BGM model:

$$\rho(T) = \rho(T)_{BGM} - C \ln T, \quad (2)$$

where $C$=19 μΩ cm is a fitting parameter. The logarithmic divergence at low temperatures indicates a possible Kondo contribution [33] below 40 K.

Supporting evidences for the Kondo picture may be found in the magnetoresistance measurement. Figure 2b shows the resistivity, $\rho$, measured for the magnetic field up to 140 KOe. Below 40 K, where the Kondo effect is thought to play a role, the resistivity becomes sensitive to the magnetic field and is gradually suppressed with increasing field, whereas it is almost field independent at higher temperatures. The inset of Fig. 2b shows the derived magnetoresistance (MR=($\rho$-$\rho_0$)/$\rho_0$×100%). We find that it is negative and reaches -7.6% at 2 K and $H = 140$ KOe. The negative MR could result from the suppression of the Kondo hybridization and consequently the reduction of the resistivity under the magnetic field. This is in strong contrast to the positive magnetoresistance observed in Y$_3$Ir$_4$Ge$_{13}$, which is caused by enhanced *s-d* band hybridization under the magnetic field [16].

Figure 3a presents the magnetic susceptibility, $\chi$, as a function of temperature measured in a field of 10 to 50 KOe parallel to the <100> direction. The ZFC and FC magnetic susceptibility data are essentially superimposable. At all fields, the susceptibility increases monotonically with decreasing temperature, showing no signs of magnetic transition. Increasing the magnetic field suppresses the susceptibility, as may be expected for the Kondo effect. The inset of Fig. 3a plots the temperature dependence of the inverse susceptibility $\chi^{-1}$. We find a good fit (the solid line) with the Curie-Weiss formula, $\chi = C/(T-\theta_p)$, above 120 K and obtain the Weiss temperature, $\theta_p = -47$ K, and an effective moment, $\mu_{eff}$ = 4.28 $\mu_B$, close to the theoretical predication of 4.54 $\mu_B$ for a free $Yb^{3+}$ ion. The large and negative value of $\theta_p$ implies strong antiferromagnetic (AFM) correlations among $Yb^{3+}$ $f$-moments. Below 40 K, a similar Curie-Weiss fit yields a reduced effective moment of ~ 3.69 $\mu_B$ possibly due to the crystal field effect.

Figure 3b shows the measured susceptibility at the magnetic field of 50 KOe parallel to the <110> and <100> directions, respectively. The two sets of data agree well at high temperatures but deviate from each other below 20 K. This anisotropy also appears in the magnetization curve at 1.6 K shown in the inset of Fig. 3b and could also be understood from the crystal field effect. At 5 T and 1.6 K, the magnetization approaches 1.31$\mu_B$ and 1.02$\mu_B$ per mol-Yb for the <110> and <100> directions, respectively. Because of the cubic symmetry, the $J$= 7/2 multiplet of the Yb $f$-orbitals is expected to split into two doublets ($\Gamma_6$ and $\Gamma_7$) and a $\Gamma_8$ quadruplet [25,34,35]. The derived effective moments at 1.6 K are consistent with the theoretical value of 1.33$\mu_B$ for a $\Gamma_6$ doublet ground state.

Figure 4a plots the specific heat $C_p$ of the $Yb_3Os_4Ge_{13}$ single crystal. At 300 K, $C_p$ is about 485.8 J mol$^{-1}$ K$^{-1}$ that approaches the Dulong-Petit limit [36], $C_v$= $3nR_0$ = 498.8 J mol$^{-1}$ K$^{-1}$, where $n$= 20 is the total number of atoms per formula unit and $R_0$ is the molar gas constant. With lowering temperature, $C_p$ decreases monotonically and shows no visible signatures of magnetic orders or structural transitions. However, as is seen in the inset of Fig. 4a, $C_p/T$ exhibits a rapid upturn below ~ 5 K.

To understand the behavior of the specific heat, we carry out a detailed analysis of the conduction electron and phonon contributions using [30]

$$C_{el+ph} = \gamma T + \frac{9R}{1-\alpha_D T}\left(\frac{T}{\Theta_D}\right)^3 \int_0^{\Theta_D/T} \frac{x^4 e^x dx}{[\exp(x)-1]^2} + \sum_{i=1}^{57} \frac{R}{1-\alpha_{Ei}T}\left(\frac{\Theta_{Ei}}{T}\right)^2 \frac{\exp(\Theta_{Ei}/T)}{[\exp(\Theta_{Ei}/T)-1]^2}, \quad (3)$$

where $\gamma T$ is the contribution of the background conduction electrons, $\Theta_D$ is the Debye temperature, $\Theta_{Ei}$ is the Einstein temperature, and $\alpha_D$ and $\alpha_E$ are the anharmonic coefficients of acoustic and optical branches, respectively. The 57 optical branches are further grouped into three inequivalent sets so the total number of the parameters is much reduced. We find the optical branches are necessary in order to fit the high temperature data. Fig. 4a presents our best fit (solid line) to the zero field specific heat and the fitting parameters are summarized in Table II. The specific heat coefficient is $\gamma$ = 14 mJ mol$^{-1}$ K$^{-2}$, consistent with our band structure calculations for the conduction electrons. We also obtain a better estimate of the Debye temperature, $\Theta_D$ =160 K.

The results for the magnetic contribution to the specific heat are presented in Fig. 4b under various magnetic fields after subtracting the background conduction electron and phonon contributions, $C_{mag}=C_p-C_{el+ph}$. In addition to the logarithmic upturn below 10 K, we find a broad peak in $C_{mag}/T$ at around 11 K that may be attributed to the Schottky anomaly associated with the excited crystal field levels. The magnetic entropy $S_{mag}$ is plotted in the inset of Fig. 4b with a constant correction for temperatures below 2 K. As is clearly seen, the magnetic entropy shows a change of slope at ~ 5 K and approaches $R_0\ln 2$ at about 10 K, suggesting a Kondo temperature of ~ 5 K for a doublet ground state.

Quantitative understanding of the magnetic specific heat can be obtained by combining the Kondo and the crystal field contributions. Using the crystal field configuration with two doublets ($\Gamma_6$ and $\Gamma_7$) and a $\Gamma_8$ quadruplet for the Yb $J=7/2$ $f$-orbitals [25,34,35], we may fit our data with an approximate analytical model [35]:

$$C_{mag} = C_{2d} - \frac{1}{k_B T^2}\left[\frac{\Delta^2 e^{-\frac{\Delta}{k_B T}}}{(1+e^{-\frac{\Delta}{k_B T}})^2}\right] + C_s, \quad (6)$$

where 
$$C_{2d} = -\frac{k_B}{2(\pi k_B T)^2}\mathrm{Re}\sum_{j=0}^{1}\left\{(\gamma_j+i\Delta_j)^2 \times\left[4\psi'\left(\frac{\gamma_j+i\Delta_j}{\pi k_B T}\right) - \psi'\left(\frac{\gamma_j+i\Delta_j}{2\pi k_B T}\right)\right]\right\} + \frac{\gamma_0+\gamma_1}{\pi T}, \quad (7)$$

and 
$$C_s = \frac{1}{k_B T^2}[(\Delta_1)^2 e^{-\frac{\Delta_1}{k_B T}} + 2(\Delta_2)^2 e^{-\frac{\Delta_2}{k_B T}} + 2(\Delta_2-\Delta_1)^2 e^{-\frac{\Delta_1+\Delta_2}{k_B T}}](1+e^{-\frac{\Delta_1}{k_B T}}+e^{-\frac{\Delta_2}{k_B T}})^{-2}. \quad (8)$$

$C_{2d}$ is the contribution of the two doublets ($\Gamma_6$ and $\Gamma_7$) including the effect of the Kondo hybridization by using finite values of $\gamma_0$ and $\gamma_1$, and $\psi'$ is the digamma function. $C_s$ is the Schottky contribution assuming three sharp crystal field levels ($\Gamma_6$, $\Gamma_7$ and $\Gamma_8$). The second term in equation (6) is introduced to correct double counting so that only one of $C_{2d}$ and $C_s$ for the two doublets contributes in the absence of hybridization ($\gamma_0=\gamma_1=0$). Physically, $\gamma_{0,1}$ represents the half-width at half-maximum of the spectral density for the two doublets and $\Delta_1$ and $\Delta_2$ are the excited crystal field energies of $\Gamma_7$ and $\Gamma_8$, respectively. For simplicity, we have neglected the broadening of the $\Gamma_8$ quadruplet, which locates at much higher energy and has little effect on the low temperature specific heat. The model contains only 4 free parameters and our best fit yields: $\gamma_0$=3.6 K for the $\Gamma_6$ doublet; $\Delta_1$=40 K and $\gamma_1$=3.3 K for the $\Gamma_7$ doublet; and $\Delta_2$ the order of 460 K for the $\Gamma_8$ quadruplet. The Kondo temperature $T_K$ at low temperatures is proportional to the width of the ground-state doublet, i.e. $\gamma_0=0.6T_K$, which gives $T_K=6$ K, consistent with our previous estimate from the magnetic entropy. The value of $\Delta_1$ is also consistent with observation of the field-induced anisotropy in the magnetic susceptibility at low temperatures. The field variation of the low temperature specific heat data may now be understood as a combination of the Kondo effect and the crystal field effect. Upon applying the magnetic field, the low temperature upturn due to the Kondo effect is suppressed and the spectral weight is pushed up to higher temperatures, causing the enhancement of the broad peak at around 10 K and the nonmonotonic temperature dependence shown in both Fig. 4a and Fig. 4b.

To see how the Kondo effect arises in $Yb_3Os_4Ge_{13}$, we have carried out the LDA band structure calculations using the refined lattice parameters listed in Table I. We take into account the spin-orbit coupling and use GGA-PBE for the exchange correlation energy with RKmax=7.0 and 1000 k-points meshes over the Brillouin zone. The sphere radii are 2.50 a.u. for Yb and Os and 2.10 a.u for Ge. The obtained band structures and density of states (DOS) are presented in Fig. 5. We see that the 14 Yb *f*-orbitals are split into the *J*=5/2 and *J*=7/2 multiplets due to the spin-orbit coupling. The *J*=7/2 multiplet of the Yb *f*-orbits locates at -0.3 eV near the Fermi energy while the *J*=5/2 multiplet is further away at about -1.6 eV. Near the Fermi energy, the LDA bands show considerable *f*-character and strong hybridization between the Yb *f*-orbitals and the Ge *p*-orbitals. This provides a microscopic support for the Kondo picture proposed from experimental analysis and can be easily understood since the Yb atoms are surrounded by the Ge-cages. The calculated total density of states at the Fermi energy is 17.1 states/eV f.u., corresponding to a specific heat coefficient of about 40.3 mJ mol$^{-1}$ K$^{-2}$. The Ge *p*-orbitals and Os *d*-orbitals contribute 5.7 states/eV f.u. or equivalently 13.4 mJ mol$^{-1}$ K$^{-2}$, consistent with the experimental $\gamma = 14$ mJ mol$^{-1}$ K$^{-2}$ contributed by the background conduction electrons.

In summary, we have successfully grown high-quality crystals of a novel 5*d* quasi-skutterudite $Yb_3Os_4Ge_{13}$ from a flux-growth method. Systematic measurements and analysis of the magnetization, resistivity, and specific heat have revealed a comprehensive picture of the crystal field configuration and the Kondo effect in the single crystal of $Yb_3Os_4Ge_{13}$. In comparison with their 3d-Co or 4d-Rh or Ru mixed-valence relatives, the 5d Yb-based germanides seem to exhibit much more intriguing properties including also the multiple charge/magnetic orders in $Yb_3Ir_4Ge_{13}$. Our observation of the Kondo behavior in $Yb_3Os_4Ge_{13}$ may bring renewed interest over the physical origin of the rich variety of physical properties in these materials. It could not be simply ascribed to the change in the chemical pressure and is beyond the conventional wisdom based on the Doniach phase diagram. Possible mechanism may involve the spin-orbit effect of the 5d elements and demands more systematic studies. Our growth of the novel quasi-skutterudite $Yb_3Os_4Ge_{13}$ single crystal may help to resolve this issue and achieve a comprehensive understanding on the role of the T site element from $3d$ to $5d$ in this or other families of materials such as $YbT_2Al_{10}$ [37].

## Acknowledgments

This research was supported in part by the Ministry of Science and Technology of China (973 Project No. 2011CBA00110 and 2011CB921701), the National Natural Science Foundation of China (No. 11274367, 11174339, and 91122034), and the Strategic Priority Research Program (B) of the Chinese Academy of Sciences (No. XDB07010100 and XDB07020200).

Table I. Crystallographic parameters of $Yb_3Os_4O_{13}$ at room temperature.[a]

| site | Wyckoff position | x | y | z | $B$ (Å$^2$) |
|---|---|---|---|---|---|
| Yb | 6d | 0.25 | 0.5 | 0.0 | 0.000(9) |
| Os | 8e | 0.25 | 0.25 | 0.25 | 0.000(5) |
| Ge1 | 2a | 0.0 | 0.0 | 0.0 | 0.000(6) |
| Ge2 | 24k | 0.0 | 0.3148(14) | 0.1496(13) | 0.010(6) |

[a] Structure parameters in space group $Pm$-$3n$ (No. 223), Z=1: $a$ = 8.9302(5) Å, $\rho_{calc}$=5.1844 g/cm$^3$ and $V$=712.17(7) Å$^3$; $R_{wp}$ = 13.87 %, $S = R_{wp}/R_e$ = 3.99, $R_F$ = 2.82%. Occupancy factors of all of the sites are unity.

Table II. Parameters obtained from the specific heat fit using Eq. (3). $\Theta_D$ is the Debye temperature, $\Theta_{Ei}$ is the Einstein temperature, $\alpha_D$ and $\alpha_{Ei}$ are the anharmonic coefficients of acoustic and optical branches, respectively.

|  | Number of branches | Characteristic temperature (K) | anharmonic coefficient (K$^{-1}$) |
|---|---|---|---|
| acoustic branches | 3 | $\Theta_D = 160$ | $\alpha_D = 5 \times 10^{-4}$ |
| optical branches | 12 | $\Theta_{E1} = 94$ | $\alpha_{E1} = 1.8 \times 10^{-4}$ |
|  | 22 | $\Theta_{E2} = 168$ | $\alpha_{E2} = 2 \times 10^{-5}$ |
|  | 23 | $\Theta_{E3} = 370$ | $\alpha_{E3} = 1 \times 10^{-5}$ |

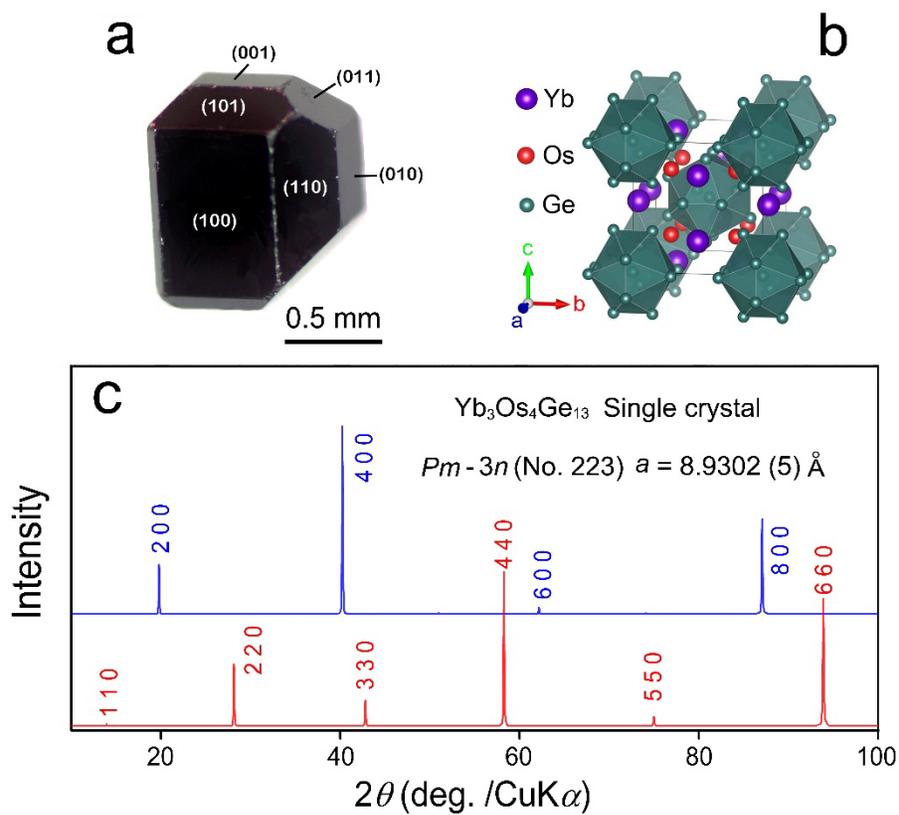

Fig. 1 (color online) a. Picture of the $Yb_3Os_4Ge_{13}$ single crystal used for this study. The large surfaces are indexed based on the XRD measurement. The size is approximately 0.8 × 0.5 × 0.5 mm$^3$; b. Crystal structure of $Yb_3Os_4Ge_{14}$ drawn based on the XRD refinement; c. XRD patterns of the crystal measured on the (100) and (110) surfaces.

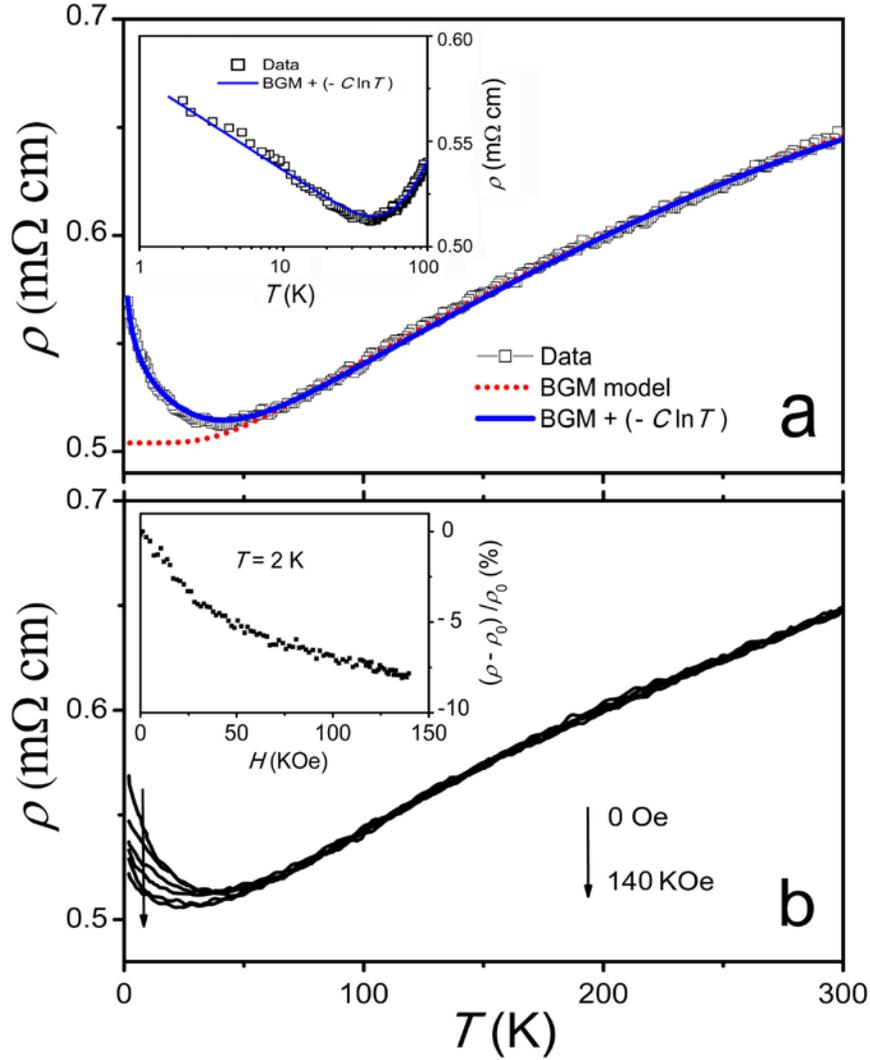

Fig. 2 (color online) a. Resistivity as a function of temperature measured in the $Yb_3Os_4Ge_{13}$ single crystal at zero magnetic field. The inset shows its logarithmic divergence at low temperatures. The dotted curve is our fit by using the Bloch-Grüneisen-Mott (BGM) model. The solid curve is our fit taking into account a logarithmic correction due to the Kondo effect in the BMG model; b. Field dependence of the resistivity, $\rho$, of the $Yb_3Os_4Ge_{13}$ single crystal. The inset shows magnetoresistance versus $H$ at 2 K.

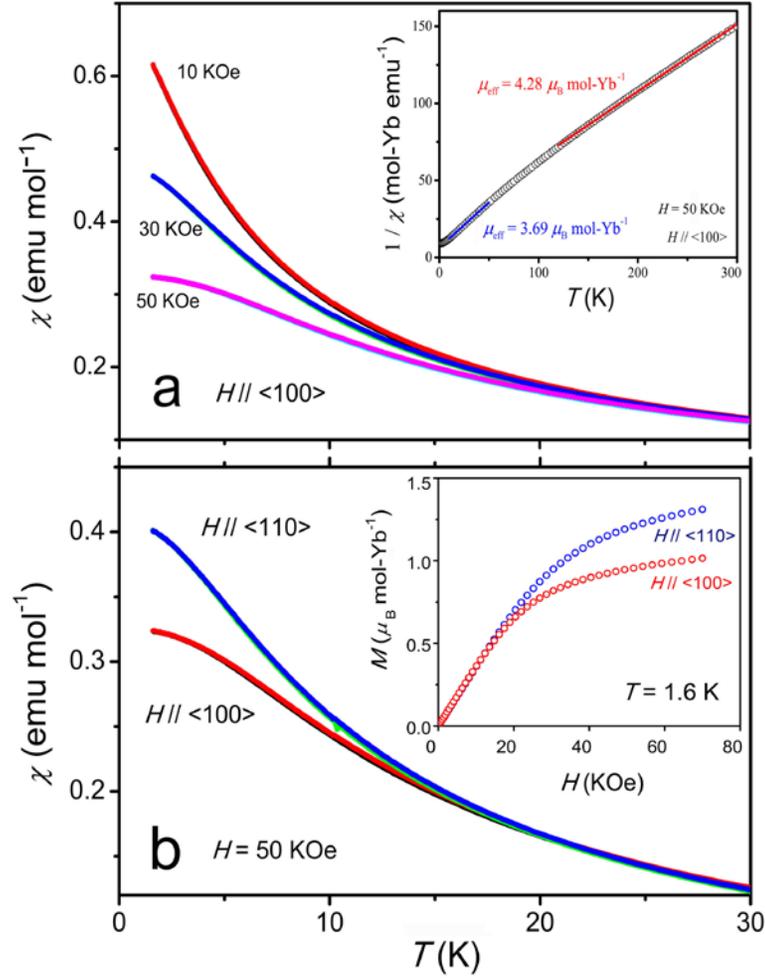

Fig. 3 (color online) The ZFC magnetic susceptibility and FC magnetic susceptibility as a function of temperature of the Yb$_3$Os$_4$Ge$_{13}$ crystal for $H$ // <100> at $H$ = 10, 30, and 50 KOe. The inset shows the Curie-Weiss fit (solid lines) to $\chi^{-1}$ in the low temperature (10-50 K) and high temperature (120-300 K) regions, respectively. b. The ZFC magnetic susceptibility and FC magnetic susceptibility for $H$ // <100> and <110>, revealing a clear anisotropy at low temperatures below 20 K. The inset shows the magnetization as a function of the magnetic field for the two directions at 1.6 K.

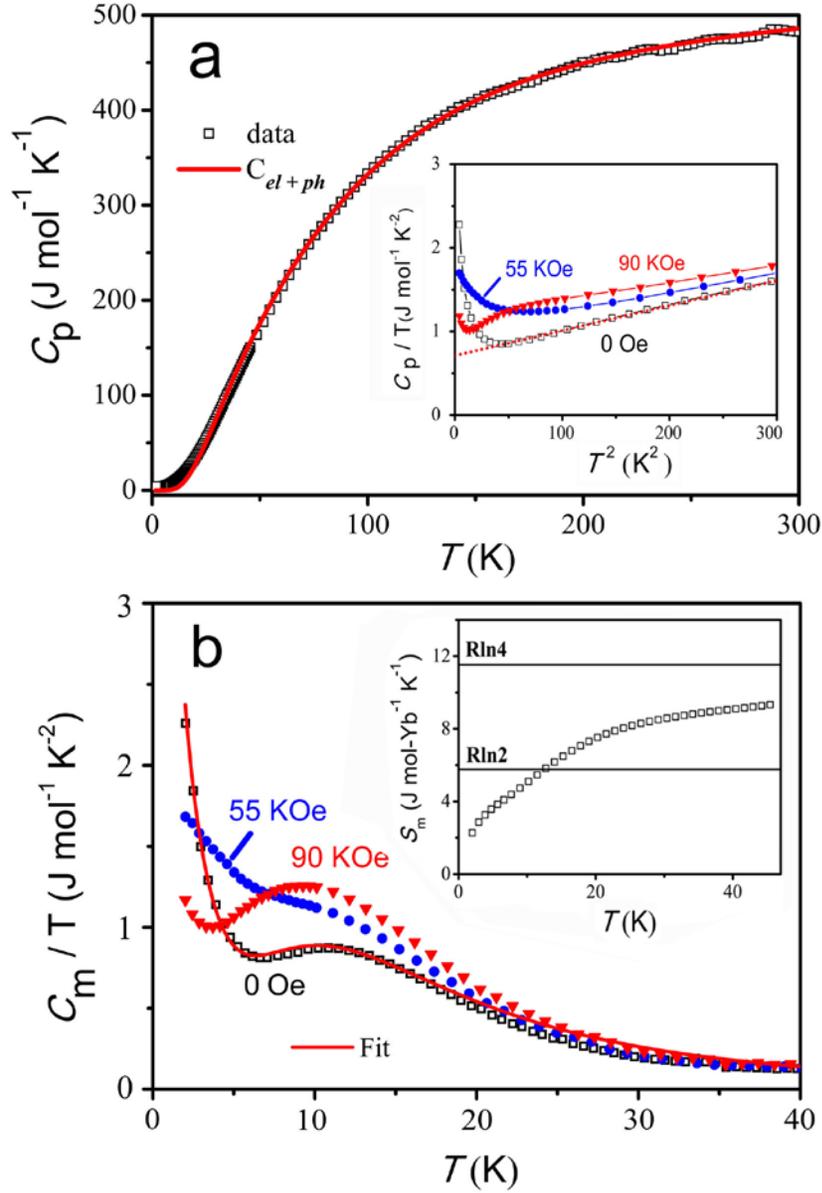

Fig. 4 a. (color online) The specific heat $C_p$ as a function of temperature $T$ (main panel) and $T^2$ (inset) of the $Yb_3Os_4Ge_{13}$ single crystal. The solid curve is the fit using the Debye-Einstein model. The dotted line in the inset is a linear fit to the low temperature data at zero field; b. The magnetic specific heat $C_{mag}$ ($C_{mag} = C_p - C_{el+ph}$) for the magnetic fields $H=$ 0, 55, 90 KOe. The solid lines are our theoretical fit taking into account both the Kondo and crystal field effects. The inset shows the magnetic entropy change as a function of temperature.

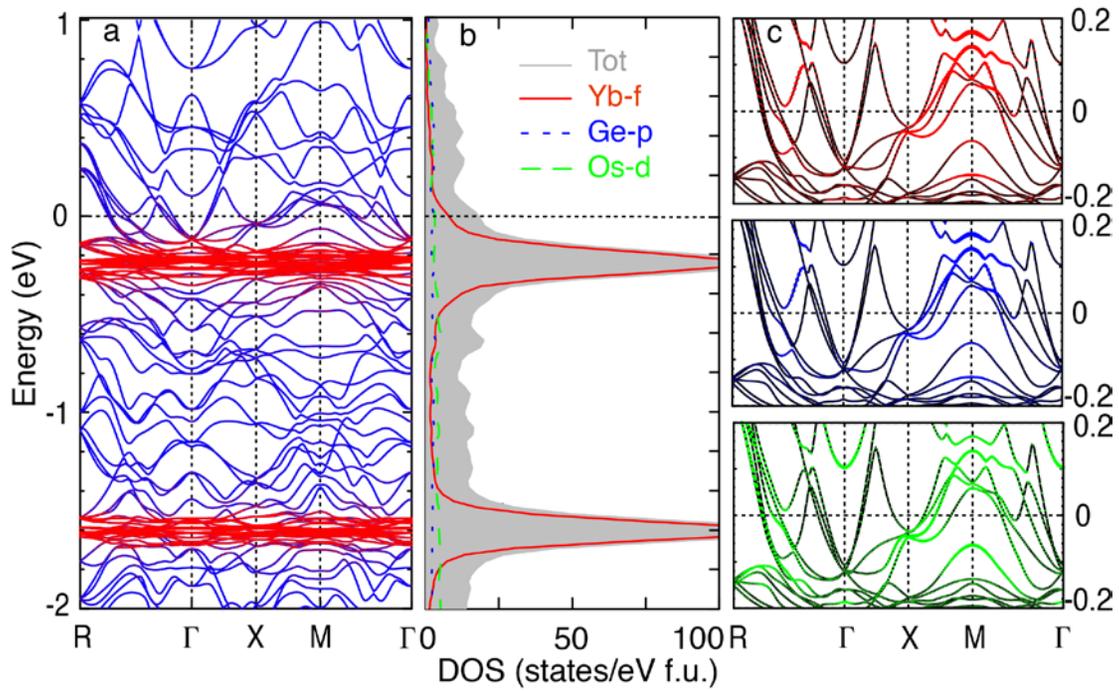

Fig. 5 (color online) The LDA calculations for $Yb_3Os_4Ge_{13}$: (a) The LDA band structures with highlighted Yb $f$-bands; (b) the total and partial densities of states; (c) Comparison of different band characters near the Fermi energy showing strong hybridization between the Yb $f$-orbitals and the Ge $p$-orbitals.